\documentclass[%
 reprint,
superscriptaddress,
nofootinbib,
 amsmath,amssymb,
 aps,
floatfix,
]{revtex4-2}

\usepackage{graphicx}
\usepackage{dcolumn}
\usepackage{bm}

\begin{document}

\preprint{APS/123-QED}

\title{12.6~dB squeezed light at 1550~nm from a bow-tie cavity for~long-term high duty~cycle~operation}
\thanks{© 2022 Optica Publishing Group. Users may use, reuse, and build upon the article, or use the article for text or data mining, so long as such uses are for non-commercial purposes and appropriate attribution is maintained. All other rights are reserved.
VoR : \url{https://doi.org/10.1364/OE.465521}
}%

\author{Biveen Shajilal}
\email{biveen.shajilal@anu.edu.au}
\affiliation{Centre for Quantum Computation and Communication Technology, Department of Quantum Science, The Australian National University, Canberra ACT 2601, Australia}
\affiliation{Centre for Quantum Computation and Communication Technology, Research School of Engineering,
The Australian National University, Canberra, ACT 2601, Australia}

\author{Oliver~Thearle}
\affiliation{Centre for Quantum Computation and Communication Technology, Department of Quantum Science, The Australian National University, Canberra ACT 2601, Australia}

\author{Aaron~Tranter}
\affiliation{Centre for Quantum Computation and Communication Technology, Department of Quantum Science, The Australian National University, Canberra ACT 2601, Australia}

\author{Yuerui~Lu}
\affiliation{Centre for Quantum Computation and Communication Technology, Research School of Engineering,
The Australian National University, Canberra, ACT 2601, Australia}

\author{Elanor~Huntington}
\affiliation{Centre for Quantum Computation and Communication Technology, Research School of Engineering,
The Australian National University, Canberra, ACT 2601, Australia}

\author{Syed~Assad}
\affiliation{Centre for Quantum Computation and Communication Technology, Department of Quantum Science, The Australian National University, Canberra ACT 2601, Australia}
\affiliation{School of Physical and Mathematical Sciences, Nanyang Technological University, Singapore 639673, Republic of Singapore}

\author{Ping~Koy~Lam}
\affiliation{Centre for Quantum Computation and Communication Technology, Department of Quantum Science, The Australian National University, Canberra ACT 2601, Australia}
\affiliation{Institute of Materials Research and Engineering, 
Agency for Science, Technology and Research (A$^\ast$STAR), 2 Fusionopolis Way, 08-03 Innovis 138634, Singapore, Republic of Singapore}
\affiliation{School of Physical and Mathematical Sciences, Nanyang Technological University, Singapore 639673, Republic of Singapore}

\author{Jiri~Janousek}
\email{jiri.janousek@anu.edu.au}
\affiliation{Centre for Quantum Computation and Communication Technology, Department of Quantum Science, The Australian National University, Canberra ACT 2601, Australia}
\affiliation{Centre for Quantum Computation and Communication Technology, Research School of Engineering,
The Australian National University, Canberra, ACT 2601, Australia}

\begin{abstract}
Squeezed states are an interesting class of quantum states that have numerous applications. This work presents the design, characterisation, and operation of a bow-tie optical parametric amplifier~(OPA) for squeezed vacuum generation. We report the high duty cycle operation and long-term stability of the system that makes it suitable for post-selection based continuous-variable quantum information protocols, cluster-state quantum computing, quantum metrology, and potentially gravitational wave detectors. Over a 50~hour continuous operation, the measured squeezing levels were greater than 10~dB with a duty cycle of 96.6\%. Alternatively, in a different mode of operation, the squeezer can also operate 10~dB below the quantum noise limit over a 12~hour period with no relocks, with an average squeezing of 11.9~dB. We also measured a maximum squeezing level of 12.6~dB at 1550~nm. This represents one of the best reported squeezing results at 1550~nm to date for a bow-tie cavity. We discuss the design aspects of the experiment that contribute to the overall stability, reliability, and longevity of the OPA, along with the automated locking schemes and different modes of operation.
\end{abstract}

\maketitle


\section{Introduction}
Squeezing has numerous applications in metrology~\cite{yonezawa2012quantum, casacio2021quantum} and secure quantum communication~\cite{madsen2012continuous,jacobsen2018complete}. The most notable is the application of squeezed light for gravitational wave sensing~\cite{tse2019quantum,aasi2013enhanced}. The development of the next-generation gravitational wave sensor, the Einstein Telescope, is progressing with 1550~nm being the wavelength of operation~\cite{collaboration2018instrument,punturo2010einstein}. The required squeezing level according to the baseline parameters is 10~dB of quantum noise suppression. Furthermore, the squeezing operation is a crucial tool in the Gaussian operation set. Continuous variable quantum information protocols that involve teleportation~\cite{furusawa1998unconditional}, cluster states~\cite{yokoyama2013ultra} and heralded gates~\cite{zhao2020high}, have squeezing as an unavoidable operation. Even though extensive research has been put into the development of squeezed state generation~\cite{30years}, the progress for a universal squeezing gate is relatively poor. Therefore, a universal squeezing gate is difficult compared to the other Gaussian operations. Conventional schemes require strongly squeezed states with very high purity that are too hard to realise experimentally~\cite{braunstein1998teleportation}.\par
J. Zhao et al.~\cite{zhao2020high} demonstrated the high-fidelity heralded quantum squeezing gate, which overcomes the limitations of traditional schemes and circumvents the strict requirements that they impose. The heralded squeezing gate managed to complete the Gaussian operation set with fidelities higher than what would be possible using non-heralded schemes, even with very high levels of ancilla squeezing. As much as the heralded scheme benefits from higher levels of squeezing for the ancilla, the protocol also benefits from the temporal stability of the squeezer. For very high target squeezing levels, the heralded scheme outperforms the conventional methods in terms of fidelity even when the ancilla squeezing is impure, however, with a lower success probability. Given the circumstance, the better temporal stability of the squeezer translates to an improved duty cycle, which translates to more data to perform the post-selection.\par 
Walshe et al.~\cite{walshe2021streamlined} used the quad-rail lattice CV cluster state to introduce improvements that allow the use of cluster state for fault-tolerant quantum computing with the GKP code using experimentally feasible squeezing levels. According to Walshe et al., the squeezing factor requirement for Clifford gates with $10^{-2}$ logical-qubit-error rates is 11.9~dB. A squeezer that could deliver these required squeezing levels in the time span over which the gates are implemented and consistently on demand, would be advantageous for cluster-state optical quantum computing. Technological advances in the field of squeezed light generation have seen important milestones, as pointed out earlier, in terms of the magnitude of quantum noise reduction~\cite{15db}. However, it is just as important to have good temporal stability for the squeezing cavity for applications like heralded squeezing gate and cluster state computing.\par
There are highly stable squeezed light sources operating at 1064~nm in gravitational wave sensors~\cite{acernese2019increasing,mehmet2020squeezed,tse2019quantum,aasi2013enhanced}. For quantum information protocols, particularly quantum communication protocols, the ideal operating window is the telecom C-band~(1550~nm). Here we present the design, characterisation and operation of a bow-tie squeezer at 1550~nm with very high duty cycles operating between 100kHz-10MHz at room temperature with no vacuum isolation. We show that while a bow-tie cavity has more optics, which may introduce more intra-cavity loss, it has the advantage of sustaining dual wavelength resonances. The resonant pump beam increases the intra-cavity power at the second harmonic. This increase in intra-cavity pump power can then be used to lower the finesse of the squeezing cavity at the fundamental and thereby increase the escape efficiency. The increase in escape efficiency in turn mitigates the effect of intra-cavity losses, consequently leading to higher measured squeezing. The bow-tie travelling wave cavity delivered maximum squeezing levels of 12.6~dB. Previously reported long-term stable cavities either operate in the air with a hemilithic configuration~\cite{acernese2019increasing,mehmet2020squeezed}, or in vacuum with a bow-tie configuration with modest levels of squeezing~\cite{tse2019quantum,aasi2013enhanced}. We report high temporal stability of our bow-tie cavity, which operates in air. This is suitable for, but not limited to, applications such as squeezing gates and cluster state quantum computing.\par
\section{Experimental setup}
\begin{figure}[b!]
\centering
\includegraphics[width=\linewidth]{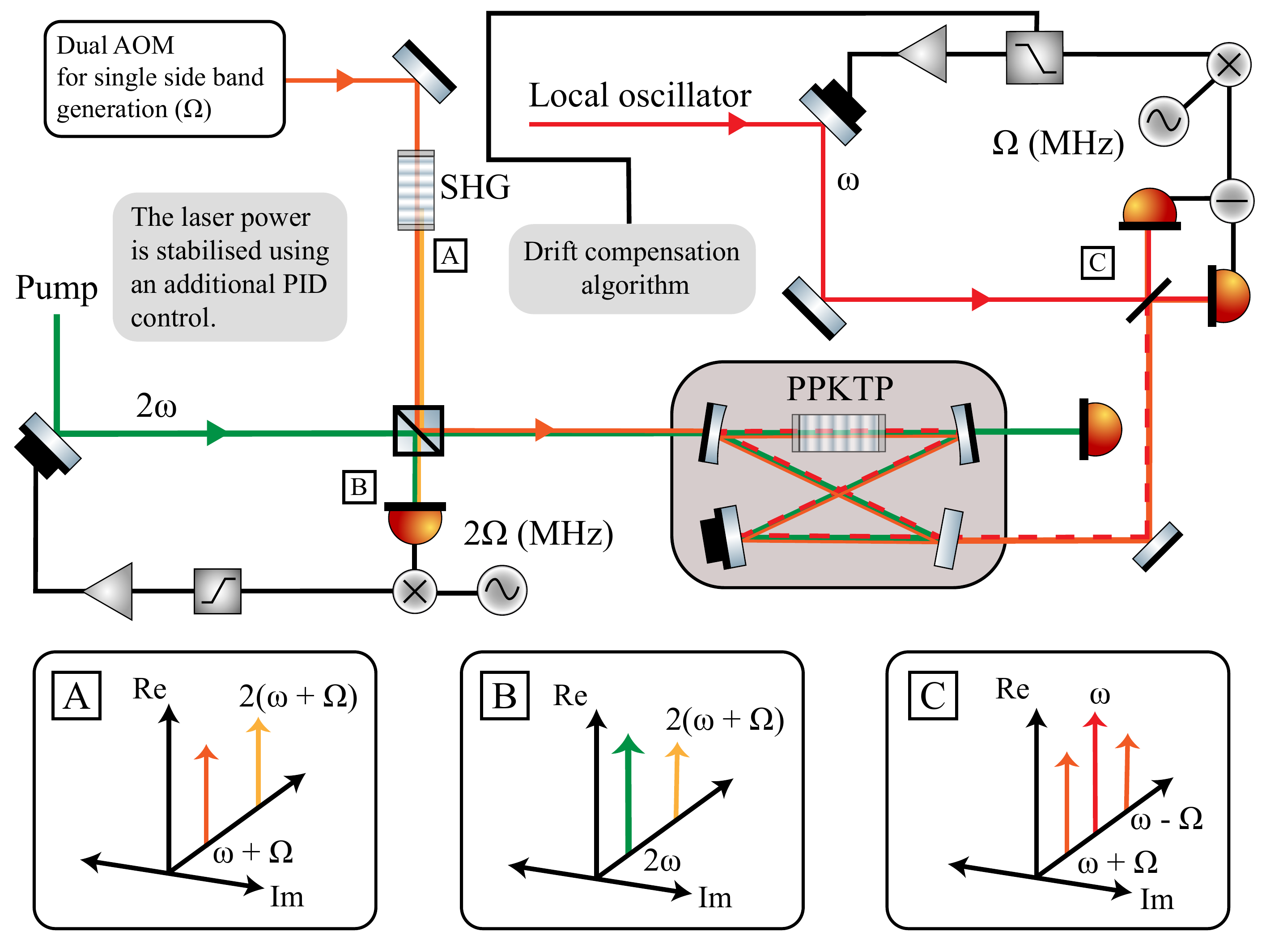}
\caption{Schematic of the optical layout of the squeezing experiment. The entire experiment comprises a bow-tie SHG cavity for the pump generation and a triangular cavity for mode cleaning of the fundamental beam. The Fig. 1(A), Fig. 1(B), and Fig. 1(C) are the phasor diagrams of the coherent control fields at A, B and C.}\label{setup}
\end{figure}

A simplified schematic of the experimental setup is presented in Fig.~\ref{setup}. The squeezed light source is a travelling wave doubly-resonant bow-tie cavity employing a  15~mm long PPKTP crystal as the nonlinear medium. The curved steering mirrors~(Radius of curvature of 50 mm), as well as the piezo-driven plane input coupler, have power reflectivity of 99.9975\% at 1550~nm. The output coupler is plane and has a reflectivity of 75\%. For the second harmonic field, the input and output couplers are high reflectivity mirrors, the steering mirrors have reflectivities of 98\% at 775~nm. A travelling wave configuration is considered inferior to a standing wave configuration, primarily because of the additional number of reflections the beam has to undergo in each round-trip. In practice, the mirrors can be near perfect and the losses due to design would not be problematic. Therefore, the dominant factor that accounts for the intracavity loss will be the non-linear crystal. PPKTP is the ideal choice because of its low loss due to scattering and absorption~\cite{ppktp}. Both faces of the nonlinear crystal are AR coated for 1550~nm and 775~nm with main emphasis to minimise loss for the fundamental. The cavity has intracavity loss of 0.1\% that corresponds to an escape efficiency of over 0.99, finesse of 21 and cavity bandwidth of 30 MHz. We tuned the laser frequency and PPKTP crystal temperature to attain double resonance within the temperature bandwidth of the PPKTP crystal. The phase matching is not optimal~(conversion efficiency is not the maximum). In addition, the crystal is slightly tilted to avoid any stray reflections that could couple into the squeezing mode. The doubly resonant operation provides reduced pump power requirements and ensures mode-matching between the pump and the squeezed mode. The pump was generated from an SHG cavity of similar configuration as the squeezing cavity.\par
For demonstrating the high temporal stability of the squeezer, it is important to avoid any phase noise by using an appropriate control scheme. The whole experiment includes several Pound-Drever-Hall~(PDH) locking loops, with the most crucial being the lock that stabilises the length of the OPA squeezing cavity. The OPA cavity being doubly resonant, we use the leakage of the pump beam in transmission through one of the curved steering mirrors to lock the OPA cavity length. This scheme avoids the usage of an additional auxiliary laser for locking the length of the cavity in the counter-propagating direction that could potentially contaminate the squeezed mode with additional technical noise through scattering/undesired reflections inside the cavity. However, to lock the relative phase of the pump field with the local oscillator field, an additional auxiliary laser is crucial to implement any coherent control technique~\cite{shapiro2000coherent}. Here, we use the modified coherent control locking scheme used by Chua \textit{et al.}~\cite{chua2011backscatter} for applications in gravitational-wave detectors. The experimental setup for the modified coherent locking is shown in Fig.~\ref{setup}. At position A, the coherent single sideband field~(CSF) shifted from the carrier~($\omega$) by $\Omega$ using dual acousto-optic modulators~(AOM) is frequency doubled with a PPKTP crystal in single pass SHG configuration. The dichroic beam splitter redirects the CSF into the OPA and the second harmonic of CSF to position B. At position B, the beat at $2\Omega$ between the upconverted CSF and the pump is measured. At position C, the beat between the CSF and the local oscillator~(LO) is measured at $\Omega$. A standard phase offset lock~\cite{shaddock2000advanced} is used to maintain a constant phase offset between the pump and CSF at B. The same scheme is used to lock CSF with LO at location C. By sequentially locking the pump to CSF and CSF to the LO, we in turn maintain a constant phase relationship between the local oscillator and the pump, that defines the angle of the squeezing ellipse. This scheme has numerous advantages over the conventional coherent locking scheme~\cite{stefszky2012generation} of which, the most important being, we decouple the pump-CSF offset lock from the OPA cavity length lock.
\section{Characterisation of the squeezing cavity}
\begin{figure}[t!]
\centering
\includegraphics[width=\linewidth]{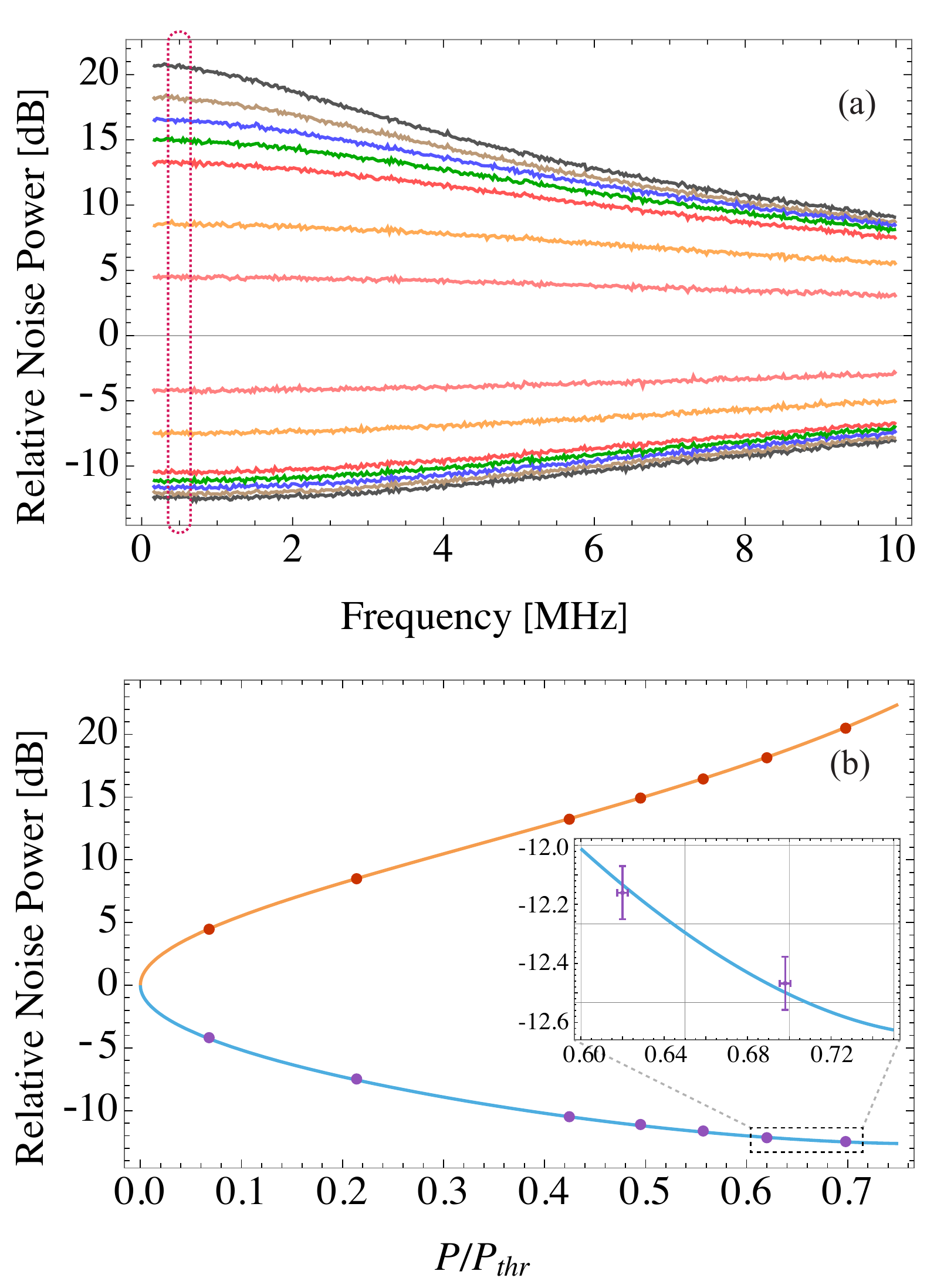}
\caption{Pump power dependence at the Fourier measurement frequency 500~kHz. The purple and orange dots are the measured values for the respective quadrature. The theoretical curves, (the blue solid and orange solid lines) were modelled using Eqs.~\ref{eqn1} \& \ref{eqn2} for phase noise $\theta_{\mathrm{jitter}}$ of 4.36~mrad and total efficiency $\eta_{\mathrm{total}}$ of $0.95\pm0.01$. The inset of Fig.~\ref{setup1}(b) shows the error on the data points over a smaller scale.}\label{setup1}
\end{figure}

For characterising the phase noise and losses in the experimental setup, the squeezing measurements were taken at different pump powers as shown in Fig.~\ref{setup1} and compared with the theoretical model for a Fourier Frequency of 500~kHz. A particular measurement window was taken to make the measurement faster by electronically filtering out the other Fourier frequency components. This helped to get a sufficiently high acquisition time, which helps to average each point in the measurement run with multiple traces in that window while maintaining a very low write off speed for the next measurement to follow. The quadrature variances for an ideal OPA with no phase noise below threshold can be written as~\cite{polzik1992atomic}, 
\begin{equation}
\Delta^2 \widehat{X}_{\pm} = 1\pm\eta_{\mathrm{total}} \frac{4 \sqrt{P / P_{\mathrm{thr}}}}{\big(1\mp\sqrt{P / P_{\mathrm{thr}}}\big)^2 + \big(2\pi f/\gamma\big)^2},\label{eqn1}
\end{equation}
where $\eta_{\mathrm{total}}$ is the total detection efficiency, $P$ is the pump power, $P_{\mathrm{thr}}=710\pm2$~mW is the pump power required to reach threshold of the OPA and $f$ is the Fourier frequency. $\gamma$ is the cavity decay rate, which is a function of losses inside the cavity. It is calculated as $\gamma = c(-\mathrm{ln}R + L)/2l,$ where $c$ is the speed of the light, $R$ is the power reflectivity of the output coupling mirror, $L$ is the round-trip loss and $l$ is the round-trip length. For our OPA cavity, $\gamma$ is calculated from measurements of $R$, $L$ and $l$ to be $9.49 \times 10^7$.\par
In addition to the net losses, it is important to consider the phase noise in our experimental setup for the theoretical model. Assuming the phase noise follows a Gaussian profile with rms value $\theta_{\mathrm{jitter}}$, this shows up as a phase jitter between the pump and LO at the homodyne detector. This will therefore couple the anti-squeezing quadrature noise into the squeezing quadrature and can be written as,
\begin{equation}
V_{\pm} = \Delta^2 \widehat{X}_{\pm} cos^2\big(\theta_{\mathrm{jitter}}\big) + \Delta^2 \widehat{X}_{\mp} sin^2\big(\theta_{\mathrm{jitter}}\big).\label{eqn2}
\end{equation}
Using Eqs.~\ref{eqn1} and \ref{eqn2}, the model was used to fit the measurements as shown in Fig.~\ref{setup1}~(b) for $f=$~500~kHz. The measurements are normalised to shot noise and corrected for electronic noise, which was around 18~dB below shot noise. The theoretical model corresponds to an $\eta_{\mathrm{total}}$ of $0.95\pm0.01$ and $\theta_{\mathrm{jitter}}$ of 4.36~mrad. The detected squeezing level is predominantly limited by the optical loss rather than the phase noise present for the given $f$. Given $\eta_{\mathrm{total}}$ = $0.95$, the residual phase noise 4.36~mrad accounts for the loss of 0.2~dB for $P/P_{\mathrm{thr}}=0.7$. Whereas, for the same amount of phase noise, improving the total efficiency by 2.5\% would increase the maximum squeezing measured by 2.8~dB at 70~\% of the threshold.
\section{Automated locking and operation modes}
As pointed out earlier, the maximum squeezing levels measured are primarily limited by the total loss rather than phase noise. This points out to how the modified coherent lock is suitable for this experiment. However, any coherent locking scheme used for this purpose has an inherent weakness. With the coherent lock, we only maintain a constant offset with the pump and LO. Any drift in pump power which could change the temperature of the non-linear medium would cause the cavity to shift the resonance. This changes the degree of squeezing. In addition, this causes a drift in the angle between the squeezed light and the local oscillator~\cite{khalaidovski2012long}. In our experiment, because of the wavelength of operation and the low absorption of PPKTP at 1550~nm, these effects are quite small. However, continuously pumping the OPA at high powers for a high degree of squeezing would mean these effects become more dominant. This drift was observed for the preliminary squeezing runs.\par
\begin{figure}[!b]
\centering
\includegraphics[width=\linewidth]{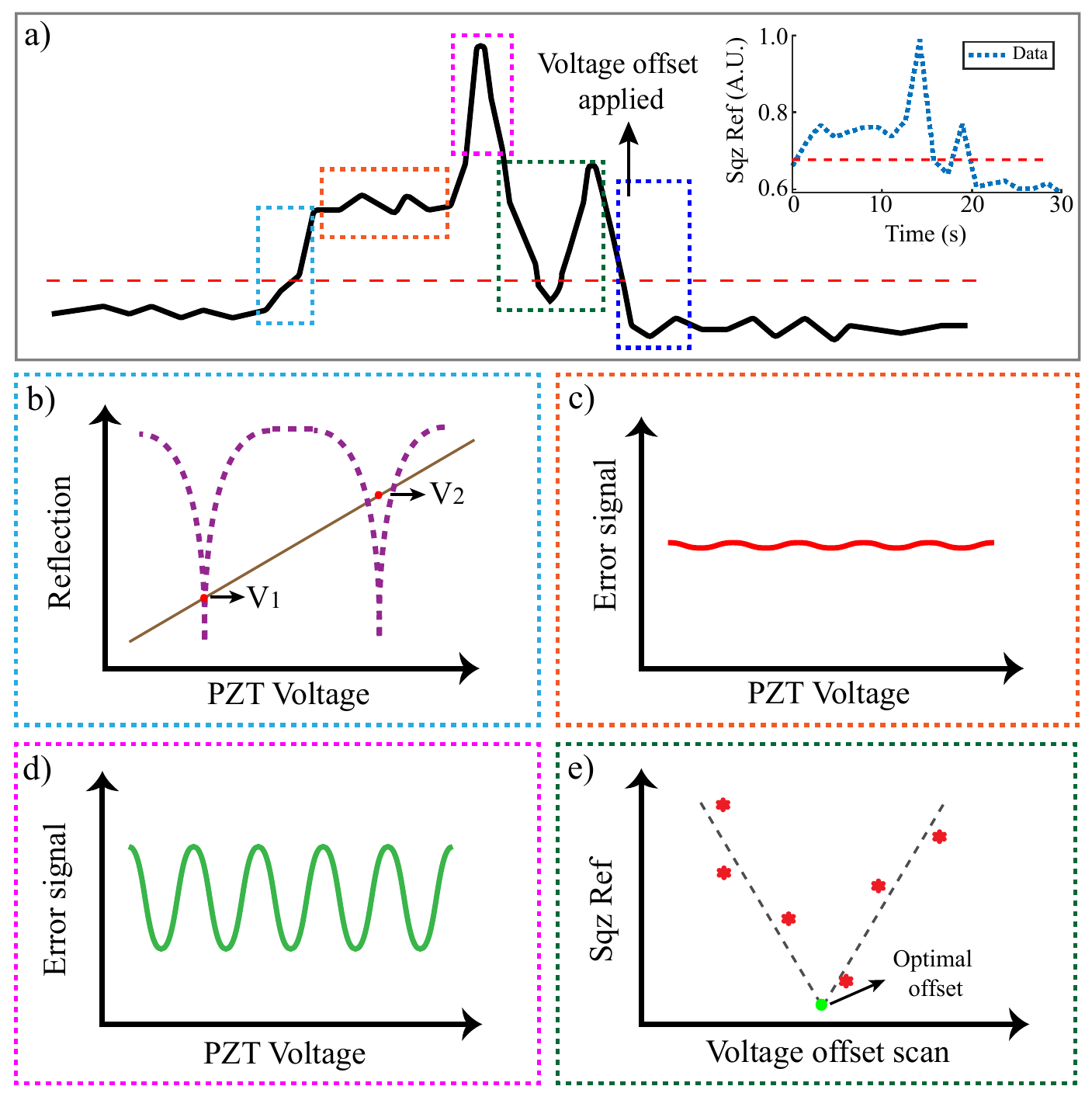}
\caption{Schematic of the Auto-relocking scheme. (a) shows the squeezing reference signal. The black trace is the squeezing reference signal~(readout of the squeezing levels) from the spectrum analyser at 500~kHz. The black trace is a schematic emphasising the features of the reference signal for a clearer understanding. The inset shows the real data on which the algorithm works. When the reference signal crosses the threshold~(9.5~dB) represented by the red dashed line, the auto-relocking sequence is initiated. The sequence starts by taking off the SHG cavity lock and OPA cavity lock along with B and C locks depicted in Fig.~\ref{setup}~i.e., DC lock between the pump and upconverted auxiliary field at 775~nm and the lock between LO and auxiliary field at 1550~nm. The locking algorithm cycles through each resonant mode in a given free spectral range~(FSR) of the OPA cavity as shown in (b), engages the pump-auxiliary field~(775~nm) lock and checks whether double resonance is achieved. Double resonance is checked by looking at the amplitude of the lock between the LO and the auxiliary field at 1550~nm. (c) shows the LO-auxiliary field when double resonance is not achieved and the lock is maximised as shown in (d) when double resonance is achieved. Once the double resonance is achieved, the optimal squeezed state-LO angle is achieved by changing the phase offset between the component locks B and C of the modified coherent lock. The squeezing values measured for a range of phase offset are measured and fitted against a v-curve to estimate the optimal phase offset, as shown in (e).}\label{algo}
\end{figure}
Further investigation showed that even though there was no degradation of the 1550~nm laser beam alignment into the SHG cavity, the laser power was drifting, which caused power drift of the pump beam as well. This affected both the modified coherent lock and the lock used to maintain the OPA cavity length. The pump power drift was occurring because of the polarisation rotation of the fibre coupled output from the 1550~nm laser. This was compensated for by the use of a local heating PID, which stabilises laser output polarisation by locally heating/cooling the polarisation maintaining fibre. Stabilising the temperature of pump power through this process helps with the drift for short intervals. Over longer intervals because of the non-vanishing absorption coefficient of the non-linear medium, local heating effects cause a drift in the squeezing angle. In addition, drifts can rise from beam pointing, acoustic vibrations and even air currents~\cite{stefszky2012generation}. Tolerance to acoustic vibrations can be further improved by using appropriate optomechanics when building the OPA cavity. The mechanical design of the cavity was tailored to avoid any such resonances with acoustic vibrations present in the lower spectrum of the audio band. For a system that is not isolated from the environment, the timescale over which this drift could occur is hard to predict/characterise as the instability of the system is a result of several effects, with some being sporadic. To ensure high duty cycle operation in such instances, it is important to implement an automated operation scheme which is fast while ensuring it locks to the mode that offers the best squeezing levels.\par
The implemented auto-relocking scheme is depicted in Fig.~\ref{algo}. The relocking scheme sequentially relocks the SHG and OPA cavities along with the two components of the modified coherent lock. The relocking mechanism is triggered when the squeezing crosses a threshold set by the user. The threshold for the experiment was set to be 9.5~dB. When the squeezing factor drops below 9.5~dB, the 4 locks are taken off lock. The SHG cavity and OPA cavities are locked in sequence, ensuring double resonance is achieved. To check this, the SHG and OPA cavities are locked in length with their respective pump beams~(1550~nm for SHG and 775~nm for OPA). The B component of the modified coherent lock is engaged, which locks the upconverted single sideband auxiliary field~(775~nm) to the pump beam. Under this condition, if double resonance is achieved, we should observe a beat between the auxiliary field~(1550~nm) and the LO. Therefore, we can use the DC error signal from C component of the modified coherent lock as the indication of double resonance. If double resonance is not achieved, we scan through the different FSR of the OPA cavity and locks to each resonant mode in a given FSR until double resonance is achieved. To make this process faster, the scanning voltage offset to the OPA-PZT~(corresponding to several resonances over a full cycle of the voltage ramp applied) are stored before every locking sequence. Once the double resonance is achieved, the last component of modified coherent lock which locks the LO to the auxiliary field at 1550~nm is engaged. To ensure the phase offset between B and C locks are optimal, the offset is scanned over a small range, the squeezing values are fitted against a v-curve and the optimal phase offset is estimated. The corresponding voltage offset to this phase offset is applied to the PZT in lock C. The whole sequence on average takes under 20 seconds to execute. Another flexibility of this approach in automating the process is that the frequency of the relock and threshold of locking can be changed based on the requirement of the experiment i.e, the cavity can operate in a situation where high levels of squeezing are required~(with a relatively higher relocking frequency) or in a case where continuous lock is required~(with a relatively lower average squeezing).\par
An alternative approach for maintaining appreciable levels of average squeezing with no relocks over longer periods is also presented. In this scheme, which we refer to as active drift compensation, the OPA locking is operated in such a way that we compensate for all the accumulated effects that would cause the squeezing angle to drift. This scheme monitors the squeezing level at the final measurement stage and changes the phase offset between the local oscillator and pump beam by tweaking the differential phase between the phase offset locks that uses the fields at positions A and B. The algorithm is a simple sequence based on the squeezing measurements at the final homodyne detector. Whenever the degree of squeezing decreases, the algorithm evaluates the gradient of the squeezing values in cycles as a response to the phase offset it applies. Once the gradient is evaluated, the algorithm gets a sense of direction of the drift relative to the previous state and estimates the polarity of the differential phase offset that should be applied. In further cycles, it compensates for the drift by applying the appropriate feedback. This offers longer continuous operation of the squeezing cavity with no relocks at higher squeezing levels.\par
\begin{figure}
\centering
\includegraphics[width=\linewidth]{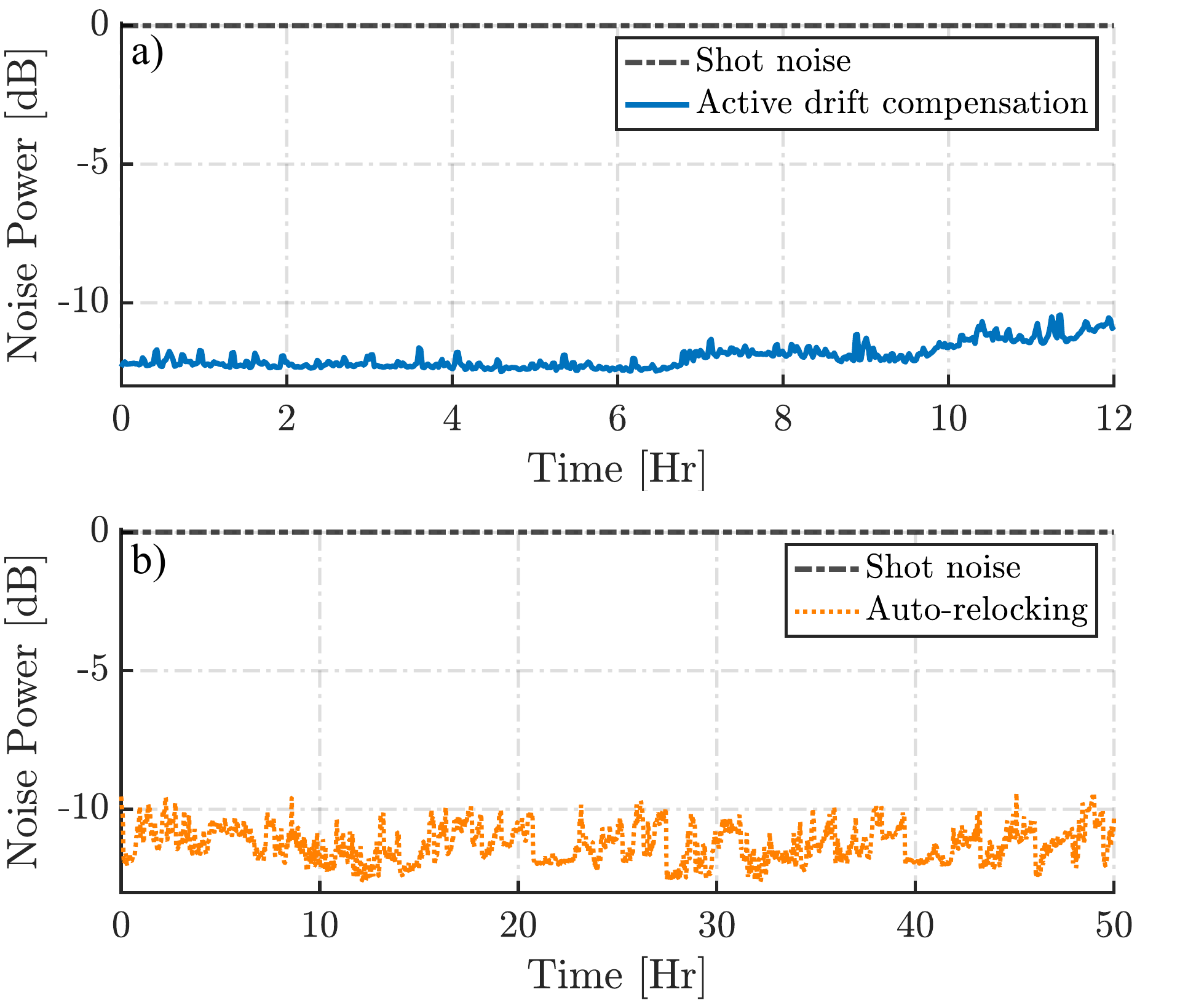}
\caption{The blue trace from (a) shows the squeezing measurement over a 12~hour window at the Fourier frequency 500~kHz where the OPA is operating in the active drift compensation mode. In this mode, an additional phase offset is actively added or removed as a response to the squeezing value readout through an automated algorithm. This ensures continuous operation that is suitable for post-selection quantum information protocols and cluster-state quantum computing. The average measured squeezing levels under this mode over the 12~hour window is 11.9~dB. The maximum measured squeezing level during this window is 12.6~dB. The orange trace from (b) represents the case where cavity is assigned to relock automatically when the squeezing drops below 9.5~dB. This operation mode is suitable for applications like gravitational wave sensing where high duty cycles are required. Over the 50~hour window, the cavity maintained high squeezing levels at the expense of additional relocks. However, the relocking times are as low as 20 seconds per lock.}\label{12hr}
\end{figure}
\begin{figure}
\centering
\includegraphics[width=\linewidth]{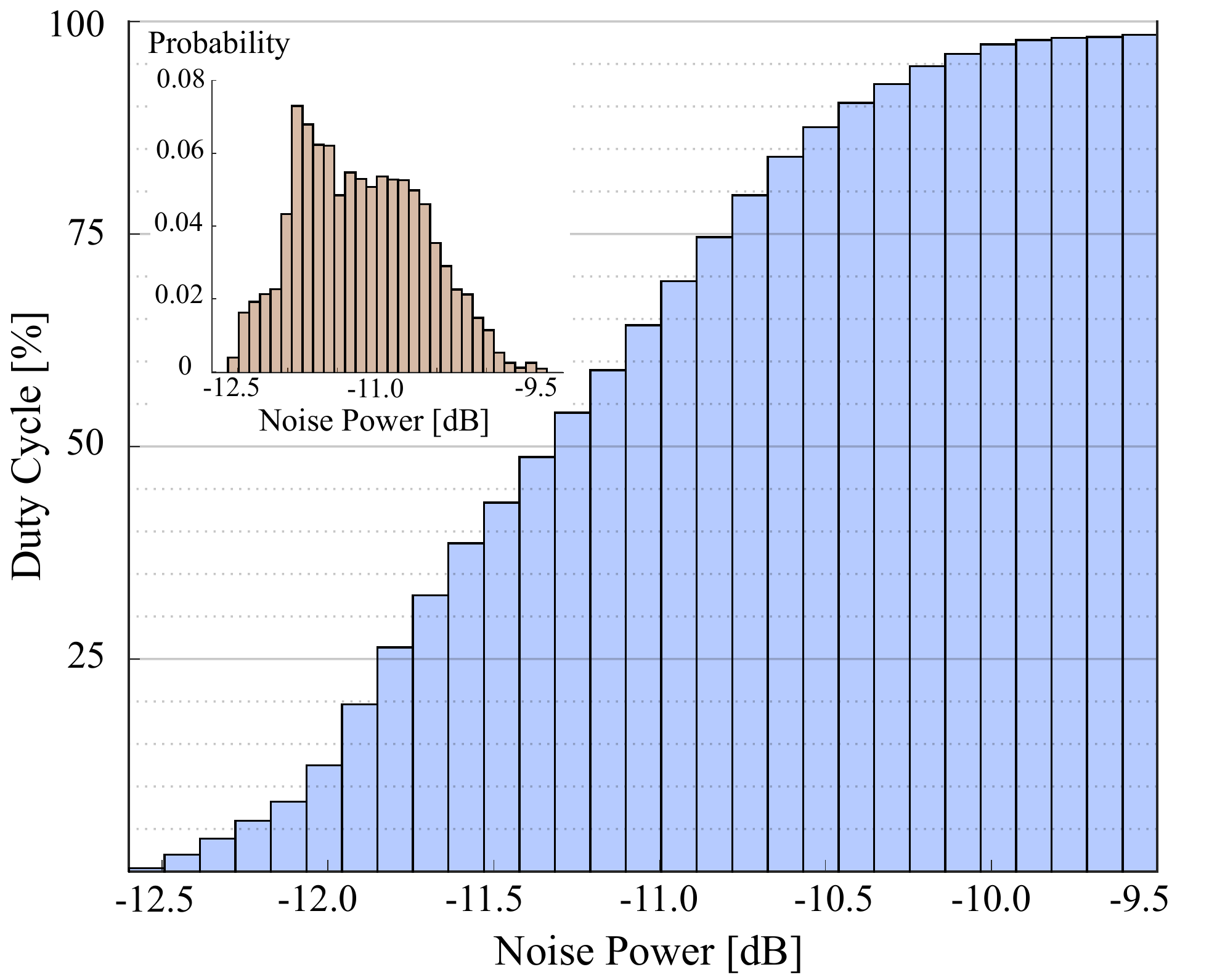}
\caption{Duty cycle of the OPA over a 50~hour window under auto-relocking mode. Each bin in the cumulative histogram represents the duty cycle of the OPA operating below the respective squeezing level. Over this 50~hour window, the OPA cavity remained locked 98.5\% of the time. The system maintained an appreciable level of squeezing below~10~dB with a duty cycle of 96.6\%. The inset shows the histogram of the squeezing measurement. The histogram peaks around 11.9~dB.}\label{heatmap}
\end{figure}
\section{Results and discussion}
The Fig.~\ref{12hr}(a) shows the temporal stability of the OPA over a 12~hour period working in the active drift compensation scheme. The squeezed light source was operated with a $P/P_{\mathrm{thr}}$ of 0.67. This condition was selected to ensure that the squeezer delivers at least 10~dB of squeezing even after the expected residual drift that could happen over 12~hours. This ensures the average squeezing stays nominally high to meet the requirements of applications pointed out earlier~\cite{walshe2021streamlined,zhao2020high}. The decrease in the measured squeezing levels over shorter time periods corresponds to how the squeezing angle is drifting, and the average drop in squeezing levels over longer periods corresponds to the slight change in resonance of the OPA~(cannot be compensated through this scheme). Over smaller timescales, when the drift occurs and crosses a certain threshold, the automated compensating sequence tries to compensate for the decrease in the measured squeezing levels by changing the phase offset between the locks that uses the fields at positions B and C. This effectively changes the angle between the squeezed mode and the LO. Ripples of alternative low and high squeezing levels occurs when the algorithm tries to apply an offset and identify the direction of drift. These are the ripples present in Fig.~\ref{12hr}(a). No relocking were required during this 12~hour window. This operation of the squeezing cavity is suitable for post-selective quantum information protocols like the heralded squeezing gate, which requires continuous operation.\par
The continuous operation of the squeezer under the active drift compensation scheme has a drawback when adapted for high duty cycle operation. The slight change in resonance of the OPA can cause a drop in the squeezing levels in continuous operation as shown in Fig.~\ref{12hr}(a) and cannot be compensated using the said scheme. The auto-relocking scheme is better suited for high duty cycle operation, as it ensures the optimal resonant mode is regained and delivers high squeezing levels. The relocking process is fast at 20 seconds to execute all the steps depicted in Fig.~\ref{algo}. Therefore, the system is ideal for experiments where the associated main setup has downtime between experimental runs and requires high levels of squeezing on demand. The lack of an active drift compensation comes at the cost of relatively more frequent relocks as the angle between the squeezed mode and LO is freely drifting from the ideal point over comparatively smaller timescales. This is shown in Fig.~\ref{12hr}(b). The squeezer was operated with a similar $P_{\mathrm{thr}}$. Under this operation, the squeezer offered an average squeezing below 10~dB during the 50~hour long run while remaining locked for 98.5\% of the time. The system maintained an appreciable degree of squeezing below 10~dB with a duty cycle of 96.6\%. Fig.~\ref{heatmap} represents the duty cycle of the OPA operating below various squeezing levels during the 50~hour window. The system also maintained an average squeezing level of 11.9~dB over 12~hours under active drift compensation as shown in Fig.~\ref{12hr}(a). This meets the squeezing requirement for implementing Clifford gates, as suggested by Walshe et al.~\cite{walshe2021streamlined}. With stability spanning a lot more than what these protocols would ideally need, these results along with the auto-relocking scheme shows the capability of our system to be widely employed in optical quantum computing platforms.\par
The highest squeezing factor reported so far at 1550~nm is 13~dB~\cite{Schonbeck18}. This experiment use a hemilithic cavity similar to the cavities used in VIRGO~\cite{acernese2019increasing,mehmet2020squeezed}. High squeezing factors~(comparable to the results from Schonbeck et al.~\cite{Schonbeck18}) and long-term stability of squeezers~(comparable to the results from LIGO~\cite{tse2019quantum,aasi2013enhanced} and VIRGO~\cite{acernese2019increasing,mehmet2020squeezed}) operating at 1550~nm in a bow-tie configuration is not reported in a single experiment. While a bow-tie cavity has more optics, which may introduce more intra-cavity loss, it also has the advantage of sustaining dual wavelength resonances. This increase in intra-cavity pump power can then be used to lower the finesse of the squeezing cavity and thereby increase the escape efficiency. The increase in escape efficiency in turn mitigates the effect of intra-cavity losses. This helps in extracting the maximum amount of squeezing from the cavity. This helped us to attain maximum measured squeezing levels of 12.6~dB at 500~kHz, which surpasses previously reported results for a bow-tie cavity and is the only experiment to combine a higher degree of squeezing and high duty cycle operation at 1550~nm.\par
\section{Conclusion}
In conclusion, we have realised a low phase noise OPA for squeezed light generation, which is capable of continuously operating squeezing levels above 10~dB for 12~hours. This is suitable for post-selective protocols, optical quantum computing, and metrology applications that require continuous operation of the squeezed light source. By comparing measured squeezing values with the theoretical model, we calculated the total efficiency as $\eta_{\mathrm{total}}=0.95\pm0.01$ and an average phase noise of 4.36~mrad over several $P/P_{\mathrm{thr}}$. The mechanical design of the OPA cavity is tolerant to acoustic resonances. This with the modified coherent lock and added phase offset compensation method provides good temporal stability, which is otherwise difficult to achieve in room-temperature/no-vacuum isolation working settings. The squeezer is capable of maintaining an average squeezing level of 11.9~dB over a 12~hour window, making it ideal for applications in optical quantum computing~\cite{walshe2021streamlined,yokoyama2013ultra}. In addition, we present an alternative operation with an automated relocking sequence. This sequence finds the optimal mode that delivers high squeezing levels over a 50~hour window with a high duty cycle of 96.6\%. The alternative operation favours usability in the context of cluster-state quantum computing and gravitational wave sensing. By having an automated operation, we ensure an immediacy to relock, and guarantee high squeezing levels. These are both important in such scenarios. The results could be extended down to the audio band with appropriate modifications to the experiment and a suitable detection scheme. This would be suitable for applications in next-generation gravitational wave detectors like the Einstein Telescope~\cite{collaboration2018instrument,punturo2010einstein}. Furthermore, by combining the two operation modes presented, the system will be capable of delivering much higher duty cycles.\par

\bibliography{apssamp}

\end{document}